# Near-Perfect Chirality and Giant Spin-Orbit Conversion in a Single Plasmonic Cavity

Lin Ma[1], Zhong-Jian Yang[1]*, Xiao-Jing Du[1], Yue You[1], Kun Zhang[2], Jun He[1]*, and Jianfang Wang[2]*

[1]*Hunan Key Laboratory of Nanophotonics and Devices, School of Physics, Central South University, Changsha 410083, China*

[2]*Department of Physics, The Chinese University of Hong Kong, Shatin, Hong Kong SAR, China*

*zjyang@csu.edu.cn; junhe@csu.edu.cn; jfwang@phy.cuhk.edu.hk

**Abstract:** To overcome the difficulty of single nanostructures in approaching the theoretical limit of chiroptical performance, we design a single plasmonic twisted dimer cavity whose magnetic gap plasmon mode enables magnetic polarization near-field engineering for high chirality. The structure exhibits strong extinction under circularly polarized excitation with one handedness, while its response to the orthogonally circularly polarized light is almost perfectly suppressed, yielding a chiral *g*-factor as high as 1.94. Meanwhile, the structure demonstrates strong chiral-selective spin-orbit angular momentum conversion: the conversion efficiency is ~95% under circularly polarized excitation with one handedness and only ~1% under the other. By tuning geometric parameters, the *g*-factor can be continuously adjusted from 0 to 1.94. Without relying on periodic coupling or collective effects, this work achieves near-perfect chirality and highly efficient angular momentum manipulation solely through intrinsic near-field matching, providing a new design strategy and theoretical basis for highly selective, ultra-compact integrated chiral photonic devices.

## Introduction

Chirality, a fundamental geometric property describing the lack of mirror symmetry, plays a crucial role in many fields, including pharmaceuticals, spintronics, quantum optics, and nanophotonics [1-5]. In particular, chiroptical phenomena are indispensable for advanced applications such as circularly polarized light detection, chiral molecular recognition, and quantum information encoding, forming the basis of key technologies like single-molecule chiral sensing, enantioselective photocatalysis, and on-chip circularly polarized light sources [6-17]. Importantly, single chiral nanostructures offer higher spatial resolution and stronger near-field confinement [18,19], making them more suitable for practical scenarios such as single-molecule detection. Enhancing their chiroptical responses is therefore critical for advancing these applications towards the real world [20,21]. Moreover, the selective manipulation of photonic spin and orbital angular momenta by chiral structures provides new degrees of freedom for quantum information encoding and light-field control [3,22,23].

The significantly enhanced light-matter interaction enabled by an extremely large chiral $g$-factor and near-perfect circular dichroism achieves circular polarization control, thereby supporting applications such as low-threshold chiral lasing and nonlinear frequency conversion[24,25]. In achieving efficient chiroptical responses, current research mainly falls into two categories: periodic architectures and individual nanostructures. Periodic structures, such as chiral metasurfaces and plasmonic arrays, can readily approach the theoretical limit of $|g| = 2$ via surface lattice resonances, quasi-bound states in the continuum, and collective coherent oscillations, emerging as the mainstream strategy for realizing giant chiral asymmetry [24-37]. In contrast, the chiroptical performance of isolated single plasmonic cavities remains severely limited. Although elaborate geometric designs have boosted the single-particle $g$-factor to ~1[38], further approaching the theoretical limit of $|g| = 2$ still faces fundamental bottlenecks. The broad linewidths of localized plasmon resonances restrict spectral selectivity. Moreover, it is extremely challenging to achieve perfect matching between electric and magnetic dipole moments and complete degeneracy breaking of chiral modes within a subwavelength volume. Combined with the absence of collective coupling effects and the intrinsic high loss of plasmonic metals, these constraints collectively hinder the further improvement of the chiroptical responses of individual nanostructures [38-58].

To address the above challenges, we propose a single plasmonic cavity based on a magnetic gap plasmon mode. By engineering its intrinsic polarization near-field, we achieve strong light absorption for one type of circular polarization while nearly perfectly suppressing the other, yielding a chiral *g*-factor of 1.94 (97% of the theoretical limit $|g| = 2$) without periodic coupling. The cavity also exhibits drastically different spin-orbit conversion efficiencies: ~95% for one handedness and only ~1% for the other, directly demonstrating selective control of spin-to-orbital angular momentum conversion. Distinct from collective designs, this work achieves near-perfect chirality and highly efficient spin-orbit conversion solely through the intrinsic mode engineering of a single cavity, offering a new strategy for miniaturized chiral photonic devices.

**Results and Discussion**

Figure 1 illustrates the physical mechanism for achieving perfect chirality. The cavity features an intrinsic polarized near-field distribution that twists helically along the propagation direction $\boldsymbol{k}$, as indicated by the orientation variation of the ellipses. Under excitation by left-handed circularly polarized (LCP, left panel), the electric-field polarization direction of the incident light matches the twisting direction of the intrinsic near-field at every cross-section, leading to in-phase superposition of the local polarization responses and thus strong near-field enhancement and absorption.

Figure 1 (right panel) illustrates the condition for achieving perfect chirality ($g = 2$) for the right-handed circularly polarized (RCP) light. The total polarization response of the structure must be coherently canceled to zero. Assuming the untwisted structure is symmetric about the central plane, the condition for a zero net response is that the induced polarization responses at symmetric positions with respect to the central plane are equal in magnitude but opposite in direction. At the central plane, the incident polarization is perpendicular to the intrinsic polarized near-field. The structural twist further creates an antisymmetric distribution of the angle between the intrinsic polarized near-field and the incident light at vertically symmetric positions, giving rise to equal-amplitude but opposite-direction excitation contributions. The polarization responses from different cross-sections cancel each other, leading to complete destructive interference of the overall response. Through this mechanism, the single resonant cavity exhibits opposite responses to left- and right-handed circularly polarized light. One experiences in-phase resonant enhancement, while the other undergoes complete coherent cancellation, thereby achieving the theoretical limit of perfect chirality ($g$ = 2).

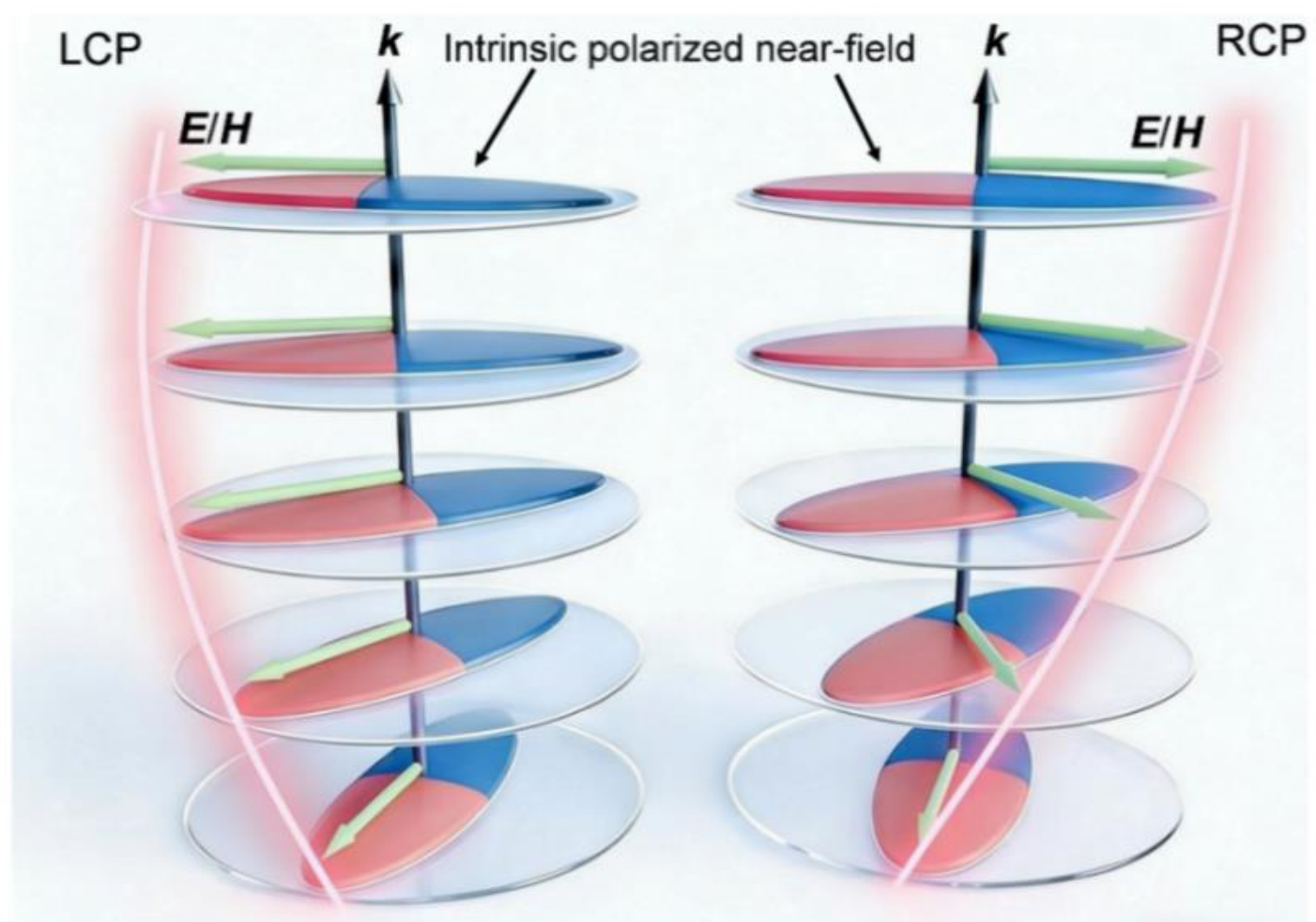

**Figure 1.** Schematic illustration of the physical mechanism for achieving perfect chiroptical responses via intrinsic polarized near-field engineering. The left and right panels show the intrinsic polarized near-field responses of a single plasmonic cavity under two orthogonally circularly polarized excitations (LCP and RCP). The ellipses represent the intrinsic polarized near-field distribution of the structure, whose orientation (color) undergoes a helical twist along the propagation direction ***k***. The green arrows denote the polarization direction (electric or magnetic field) of the incident light.

In a practical structural design, the primary prerequisite for realizing perfect chirality is to ensure high mode purity near the operation frequency. Based on this principle, we propose a twisted dimer structure composed of two metallic nanorods separated by a narrow gap to support magnetic gap plasmon modes. The entire structure is twisted by a certain angle along the light propagation direction, as illustrated in Figure 2(a). Numerical simulations were performed using a finite-difference time-domain (FDTD) method (see Method in the Supporting

Information). Although the magnetic dipole near-field is concentrated in the central region and weakened at both ends, the integrals of the near-field responses over different cross-sections exhibit a nearly symmetric distribution with respect to the central plane (Fig. S1). Accordingly, the aforementioned design strategy of coherent cancellation remains valid for this structure.

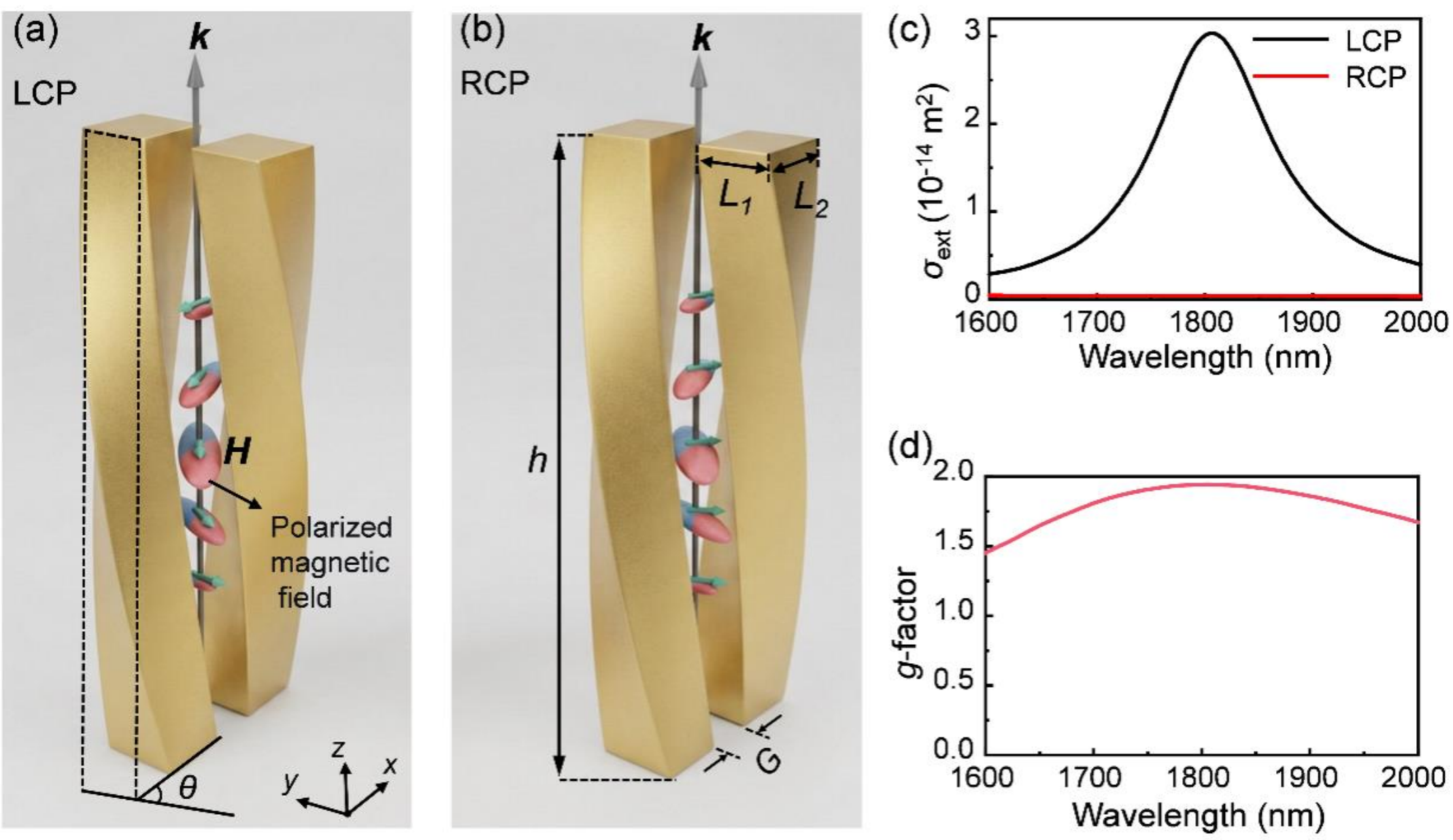

**Figure 2.** Designed plasmonic cavity structure and numerical simulation results. (a,b) Schematics of the twisted plasmonic cavity under LCP (a) and RCP (b) excitation. The ellipses represent the intrinsic magnetic near-field responses in the gap, with their sizes indicating the local field intensity. The light green arrows denote the magnetic field polarization of the incident light. Geometric parameters: height $h$, twist angle $\theta$, gap between the two arms $G$, arm lengths $L_1$ and $L_2$. (c,d) Corresponding extinction cross-section spectra (c) and chiral $g$-factor spectrum (d). Simulation parameters: $h$ = 350 nm, $\theta$ = 70°, $G$ = 10 nm, $L_1$ = $L_2$ = 50 nm. The $g$-factor reaches 1.94 at the resonance wavelength.

According to the relationship between the structural height $h$ and the resonance wavelength $\lambda$ in Figure 2, the cumulative polarization rotation of circularly polarized light passing through the entire structure is given by

$$\theta = (h / \lambda) \times 360° \quad (1)$$

which yields $\theta$ = 70°. We therefore set the geometric twist angle to 70° to precisely match the polarization rotation of light (Figure 2a and 2b). The simulation results show that under LCP excitation, the structure exhibits strong extinction dominated by absorption (see Figure S1), while under RCP excitation, the extinction is almost completely suppressed (Figure 2c). The *g*-factor reaches 1.94 (Figure 2d), very close to the theoretical limit $|g| = 2$. In this system, the intrinsic magnetic polarization direction rotates uniformly along the height, following the same rotation as the LCP polarization. Consequently, the LCP polarization remains essentially parallel to the local magnetic polarization at all heights, leading to in-phase superposition of the responses from different layers (Figure 2a). In contrast, the RCP polarization is approximately perpendicular to the magnetic polarization at the central plane, and the contributions from vertically symmetric positions cancel each other (Figure 2b).

It is worth emphasizing that for a fixed height *h*, the *g*-factor remains nearly unchanged ($g > 1.92$) within a certain range ($\pm 10°$) around the matched twist angle, demonstrating excellent angular robustness of the proposed design. The response under larger angular deviations will be further discussed below. Meanwhile, we applied the same design concept to a twisted planar structure to achieve a chiral response dominated by the electric dipole mode. Although this configuration yields a *g*-factor of ~1.2 and fails to approach the theoretical limit due to interference from neighboring higher-order modes (Figure S2), it still exceeds the values reported for the vast majority of single-unit chiral structures, further highlighting the efficiency of our design strategy.

The near-perfect chiral response of the twisted dimer described above can also be understood from the perfect matching between the total electric dipole moment ($\boldsymbol{p}$) and the total magnetic dipole moment ($\boldsymbol{m}$) in its intrinsic near-field response. The magnetic dipole moment $\boldsymbol{m}$ is aligned with the magnetic field at the central plane of the structure, forming an angle of 125° with respect to the *x*-axis (Figure 2). The electric dipole moment $\boldsymbol{p}$ mainly originates from the superposition of the electric fields around the two ends of the dimer. Under the structural twist, its effective direction makes an angle of about 55° with the *x*-axis, antiparallel to $\boldsymbol{m}$. Moreover, $\boldsymbol{p}$ and $\boldsymbol{m}$ exhibit a phase difference of $\pi/2$ (Figure S3), resulting in $|\mathrm{Im}[\boldsymbol{p}\cdot\boldsymbol{m}^*]| \approx 1$. Consequently, this leads to the near-perfect chiral response.

It should also be emphasized that the excitation of the entire system is governed by magnetic polarization near-field matching (an intrinsic characteristic of the magnetic-dipole-dominated mode). As a passive component induced by the magnetic mode response, which can be verified by changing the excitation scheme (Figure S4), the electric dipole response participates only in chiral coupling and radiation manipulation, without altering the physical origin of the chiral selectivity. Furthermore, despite the near-perfect chirality, the far-field radiation of the electric and magnetic dipoles undergoes destructive interference. This suppresses scattering and ensures that absorption dominates the extinction, even though both dipoles exhibit strong near-field responses.

We further verified the underlying mechanism of the near-perfect chirality in this system, the design principle based on magnetic polarization near-field matching, through geometric parameter tuning. Figure 3a presents the variation of the *g*-factor with the twist angle $\theta$ under a given $h$. Within a range of $\pm 10°$ around the optimal twist angle (70°), the *g*-factor remains close to the theoretical limit. As the twist angle deviates further from the optimum, the *g*-factor decreases. This is due to phase retardation under plane-wave excitation, which breaks the equal-magnitude /opposite-direction symmetry of the local responses about the central plane. Larger deviations from the matched angle amplify this asymmetry, thus weakening the destructive interference for RCP light and lowering the *g*-factor (Figure S5). On the other hand, when the structural height $h$ is varied, as long as the mode purity is maintained and the twist-angle matching condition eq 1 is satisfied, near-perfect chiral responses are obtained (Figure 3b). When $h$ becomes too low, the *g*-factor drops because the tails of other modes enter the working wavelength range and interfere with the purity of the main mode (Figure S6).

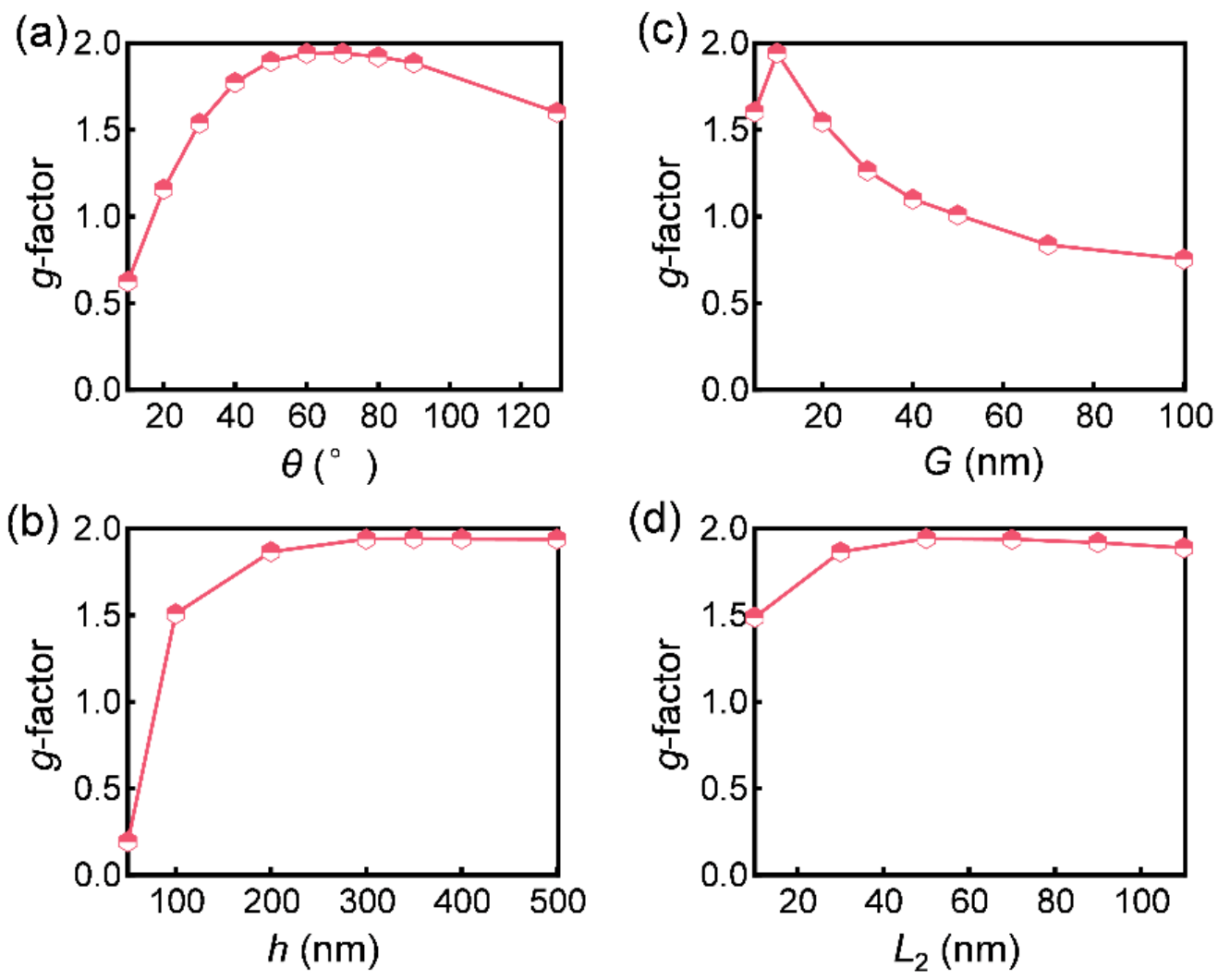

**Figure 3.** Influence of the geometric parameters on the chiral response of the cavity. Keeping all other structural parameters the same as in Figure 2, the variation of the *g*-factor with the individual geometrical parameters were investigated. (a) Twist angle $\theta$. (b) Height $h$. (c) Dimer gap $G$. (d) Arm length $L_2$.

The dimer gap $G$ affects the *g*-factor by modifying the mode purity and retardation effect. Reducing the gap improves the purity of the magnetic gap plasmon mode and significantly weakens the phase-retardation effect, leading to a corresponding increase in the *g*-factor. However, when the gap becomes too small (e.g.,~5 nm), although the mode becomes purer, the intrinsic magnetic dipole response strength drops markedly, causing the *g*-factor to decrease instead (Figure S7). Similarly, as long as the magnetic gap plasmon mode remains pure and the twist-angle matching condition is satisfied (Figure S8), the system maintains near-perfect chiral responses even when only the arm length $L_2$ is varied (Figure 3d). These results further confirm that the chiral selectivity is determined by the magnetic polarization near-field matching of the magnetic gap plasmon mode.

We further analyzed the evolution of the orbital angular momentum (OAM) along the propagation direction $z$ (Figure 4). The longitudinal OAM at a specific cross-section, denoted as $L_z$, was calculated by integrating the $z$-component of the OAM density, over a defined transverse area $A_l$, $L_z = \int_{A_l} l_z \, dA$. To accurately capture the localized near-field distribution surrounding the nanostructure, the integration area $A_l$, was chosen to be slightly larger than the physical cross-section of the structure (Figure S9). The other orthogonal components of the OAM are negligible. The time-averaged OAM density vector $l$ is defined as $l = \frac{\mathrm{Im}[\varepsilon(E^{*}\cdot(r\times\nabla)E)+\mu(H^{*}\cdot(r\times\nabla)H)]}{4\omega}$, where $\varepsilon$ and $\mu$ are the permittivity and permeability of the medium, and $*$ denotes complex conjugation. To evaluate the overall OAM, we averaged $L_z$ along the propagation direction $z$ to obtain the mean value $\langle L_z\rangle$. For the conversion efficiency normalization, the incident spin angular momentum (SAM $\langle S_z\rangle$, with other components being negligible), was calculated in the absence of the structure. This was performed by integrating the $z$-component of the spin density $s_z$ over an effective area $A_s$, $\langle Sz\rangle = \int_{A_s} s_z\, dA$, where the spin density vector $s = \frac{\mathrm{Im}[\varepsilon(E^{*}\times E)+\mu(H^{*}\times H)]}{4\omega}$, The integration area $A_s$ was determined by the equivalent extinction cross-section of the structure (i.e., the average of the extinction cross-sections under LCP and RCP excitations).Under LCP excitation, the incident SAM is efficiently converted into OAM via spin-orbit conversion (SOC), yielding an average conversion efficiency of $\eta_{\mathrm{LCP}} = |\langle L_z\rangle|/|\langle S_z\rangle| \approx 95\%$. Under RCP excitation, the OAM signal almost completely vanishes, with an efficiency $\eta_{\mathrm{RCP}} \approx 1\%$.

Notably, the SOC efficiency can be substituted into the unified definition consistent with the chiral $g$-factor, expressed as $g_{\mathrm{SOC}} = 2(\eta_{\mathrm{LCP}} - \eta_{\mathrm{RCP}})/(\eta_{\mathrm{LCP}} + \eta_{\mathrm{RCP}})$. The calculated $g_{\mathrm{SOC}}$ reaches ~1.94, which is in excellent agreement with the previous chiral $g$-factor. This result reveals an approximate “all-pass/all-block” effect of OAM conversion enabled by magnetic polarization near-field matching. For LCP light well matched with the chiral near-field, the system exhibits an all-pass response, where efficient SOC is realized, corresponding to the maximal chiral absorption. By contrast, for RCP light mismatched with the chiral near-field, the system presents an all-block behavior, where OAM generation is strongly suppressed and the chiral absorption reaches its minimum. Such extreme selectivity of the angular momentum conversion channel demonstrates the intrinsic physical unity between chiral absorption and angular momentum manipulation. Both phenomena are fundamentally governed by magnetic polarization near-field matching, further verifying the dominant role of this mechanism in the twisted plasmonic dimer.

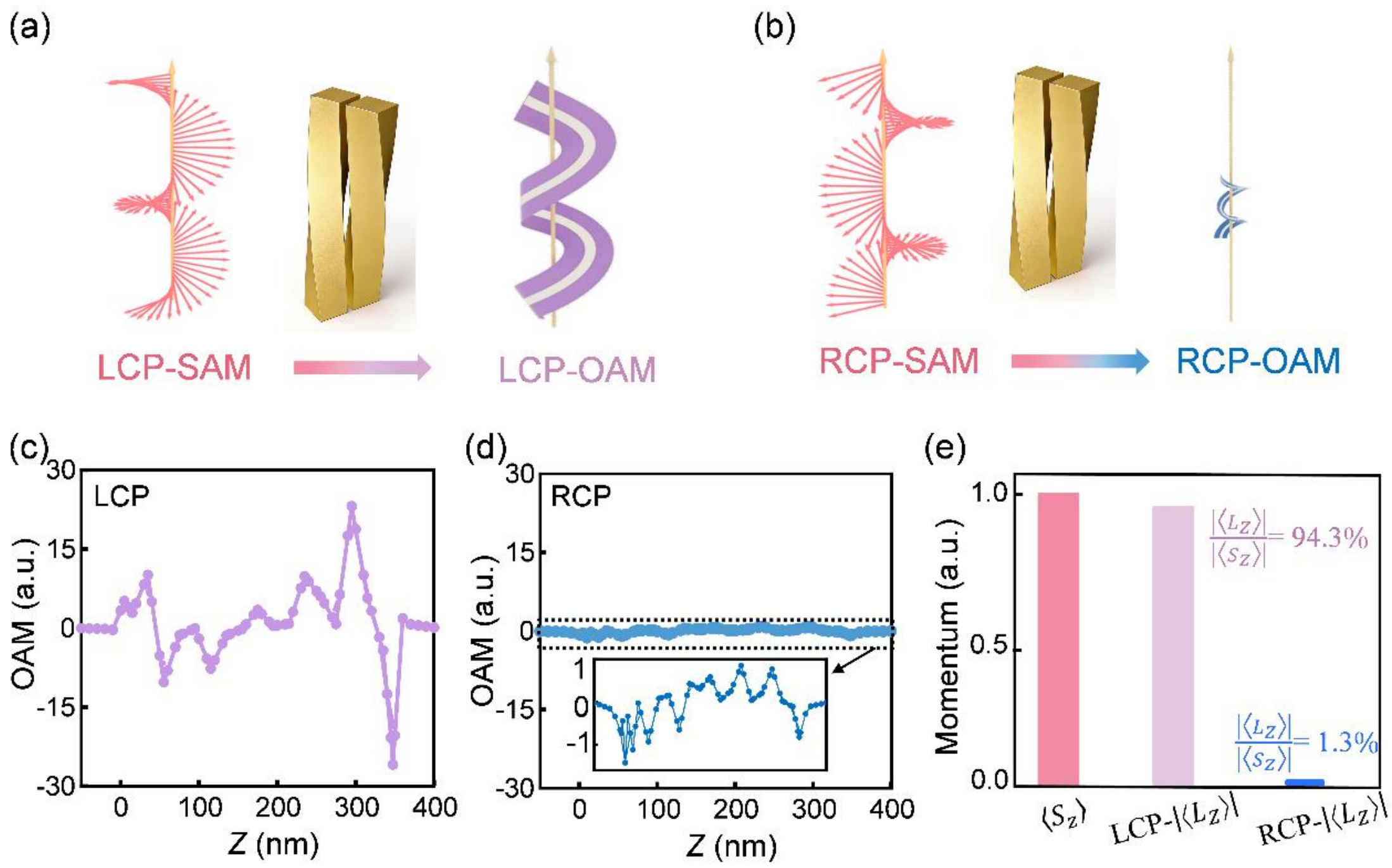

**Figure 4.** Chiral-selective SOC. (a,b) Schematic illustrations of near-field SOC under LCP (a) and RCP (b) excitation. (c,d) Evolutions of OAM along the propagation direction: pronounced OAM under LCP excitation (c); suppressed OAM under RCP excitation, with an inset showing the magnified view (d). (e) SOC efficiency: $|\langle L_z\rangle|/|\langle S_z\rangle|$= 94.3% under LCP and 1.3% under RCP.

## Conclusion

In conclusion, in this work we have designed a single plasmonic cavity based on a magnetic gap plasmon mode. By constructing a pure magnetic dipole mode and precisely controlling the matching between its intrinsic magnetic polarization near-field and the excitation polarization field, significant absorption is achieved under circularly polarized light with one handedness while the response to the orthogonal circular polarization is nearly completely suppressed. Without introducing periodic coupling or collective effects, the system attains a near-theoretical-limit chiral response ($g$ = 1.94) and exhibits excellent chiral-selective angular momentum manipulation, with the SOC efficiency being ~95% under LCP excitation and only ~1% under RCP excitation. This work overcomes the long-standing bottleneck that the chiral response of single plasmonic structures hardly approaches the theoretical limit, demonstrating that near-perfect chirality can be realized solely through intrinsic mode matching. It provides new design concepts and theoretical foundations for ultra-compact chiral photonic devices and high-performance angular momentum manipulation devices.

**Acknowledgements**

This paper was supported by the National Natural Science Foundation of China (Grant No. 11704416) and the Research Grants Council of Hong Kong (GRF, 14303124).

# Supporting Information for

# Near-Perfect Chirality and Giant Spin-Orbit Conversion in a Single Plasmonic Cavity

Lin Ma[1], Zhong-Jian Yang[1]*, Xiao-Jing Du[1], Yue You[1], Kun Zhang[2], Jun He[1]*, and Jianfang Wang[2]*

[1]*Hunan Key Laboratory of Nanophotonics and Devices, School of Physics, Central South University, Changsha 410083, China*

[2]*Department of Physics, The Chinese University of Hong Kong, Shatin, Hong Kong SAR, China*

*zjyang@csu.edu.cn; junhe@csu.edu.cn; jfwang@phy.cuhk.edu.hk

## 1. Method

The absorption/scattering/extinction cross-sections, and near-field distributions of the electric and magnetic fields were calculated using Finite-Difference Time-Domain (FDTD) Solutions 2020 R2.4 (Lumerical). The mesh size around the coupled metallic helical structure in the simulations was $3 \times 2 \times 1$ nm$^3$. In the FDTD simulations, the circularly polarized plane waves (LCP and RCP) were created by superposing two linearly polarized plane waves with a phase difference of $\pm 90°$ (90° for LCP and −90° for RCP) and used as the excitation light. The refractive index of Au was taken from Palik's book [1]. Perfectly matched layers were set in the *x*, *y*, and *z* directions.

## 2. Supporting Figures

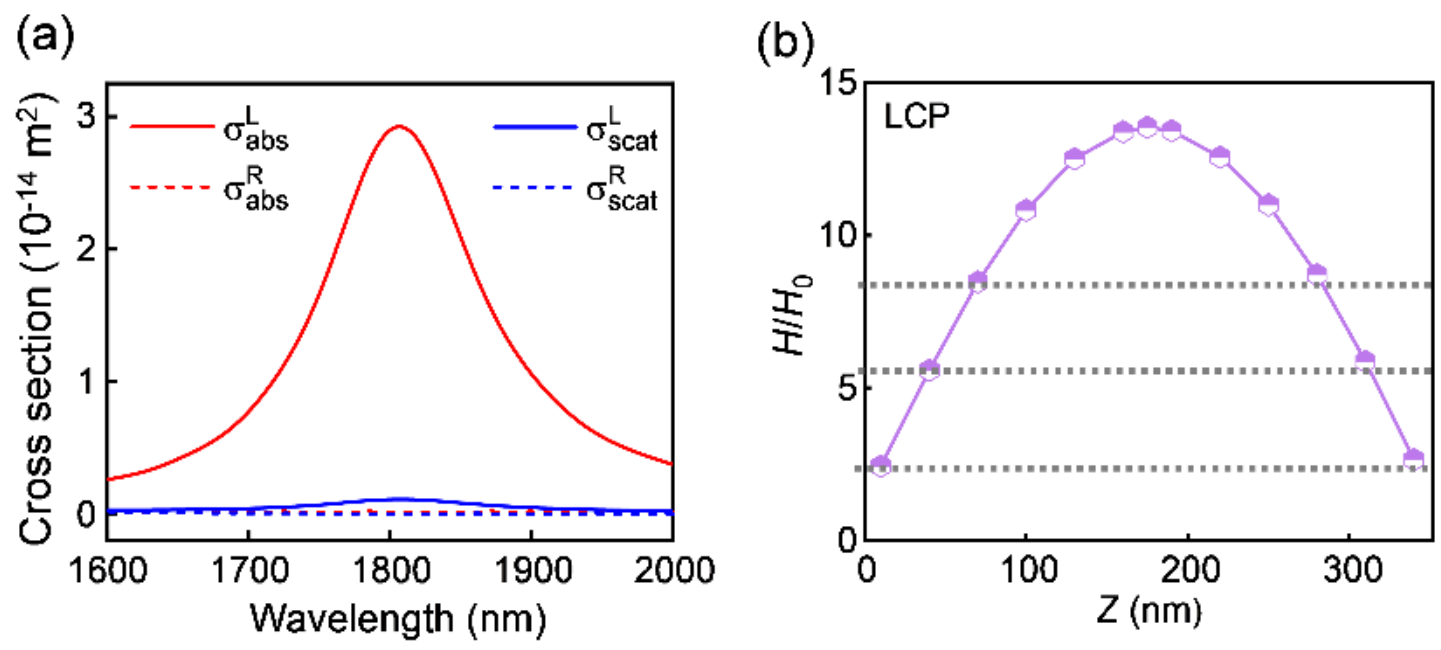

**Figure. S1** Optical and magnetic responses of a twisted dimer structure under CP excitation. (a) Absorption (red) and scattering (blue) cross-sections under LCP (solid curves) and RCP (dashed curves) excitation. (b) Magnetic field enhancement at the center of a twisted dimer structure as a function of height *Z*. The gray dashed line indicates that the magnetic field enhancements at two heights symmetric about the central height $Z$ = 175 nm (e.g., $Z$ = 10 nm and $Z$ = 340 nm) are almost equal.

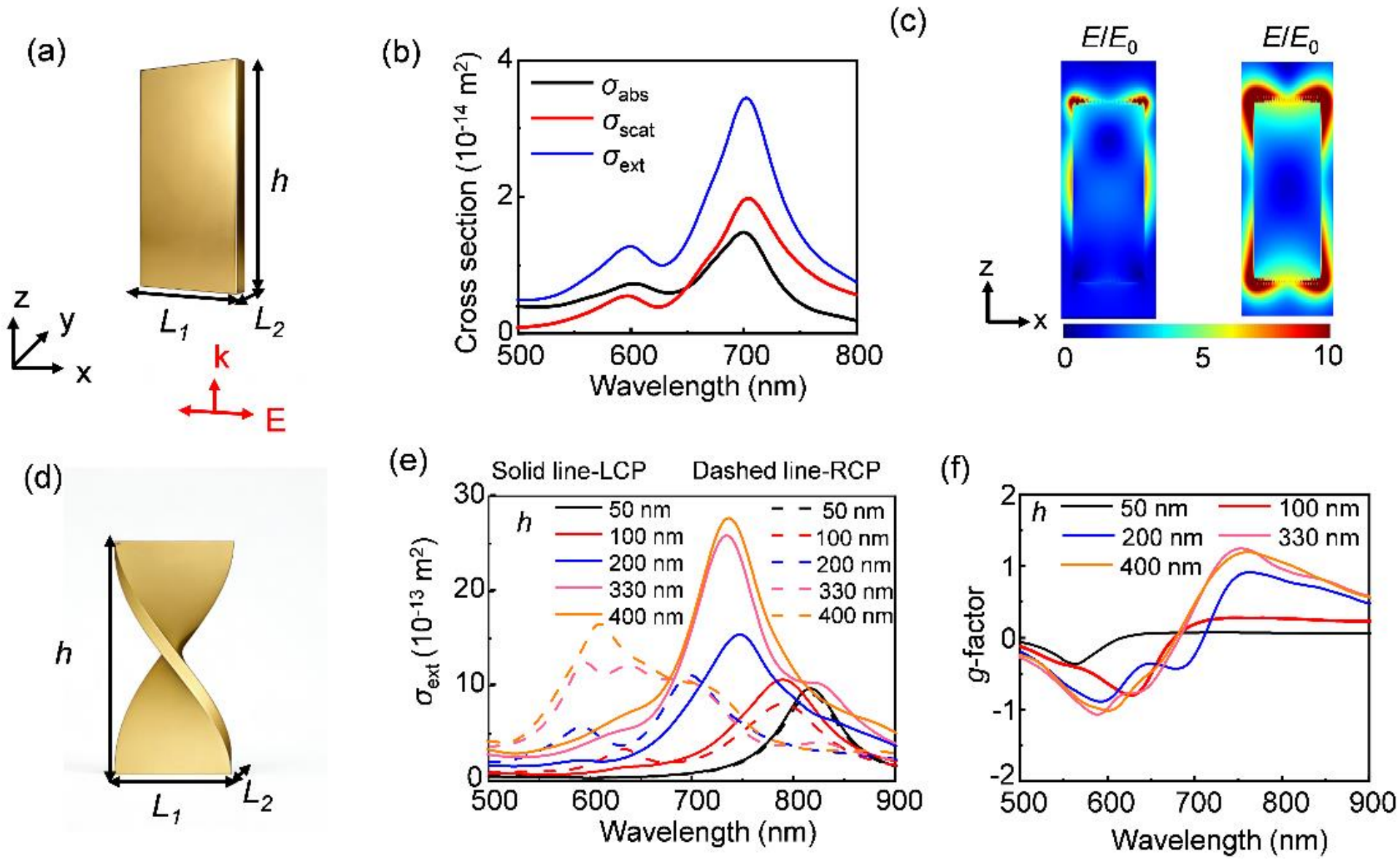

**Figure S2** Optical properties of a nanorod and a twisted nanorod structure for different heights. (a) Schematic diagram of a nanorod under *x*-polarized plane wave excitation. The length, width, and height of each nanorod are $L_1$ = 150 nm $L_2$ = 20 nm, $h$ = 350 nm. (b) Absorption (black), scattering (red), and extinction (blue) cross-sections of the nanorod. (c) Electric field distributions at the resonance wavelengths of 598 nm (left) and 702 nm (right), corresponding to the higher-order mode and the electric dipole mode, respectively. (d) Schematic diagram of the nanorod with a twist angle of 180°. (e) Extinction spectra under LCP (solid curves) and RCP (dashed curves) excitation for $h$ = 50 nm，100 nm, 200 nm, 330 nm, and 400 nm. (f) Corresponding *g*-factor spectra for $h$ = 50 nm, 100 nm, 200 nm, 330 nm, and 400 nm.

We also designed a twisted chiral nanorod that was dominated by an electric dipole mode. A single slender rod with a large aspect ratio can provide a pure and strong electric dipole resonance (Figure S2a–c), where the refractive index of gold was taken from Palik [1], and the surrounding medium had a refractive index of 1. An electric dipole mode appears at a resonance wavelength of 702 nm, with its electric field distribution shown in Figure S2c. The gold rod is helically twisted, with the twist angle given by $\theta = (h/\lambda) \times 360°$ (Figure S2d). Figure S2e presents the simulated extinction spectra for LCP and RCP excitation at heights $h$ = 50 nm, 100 nm, 200 nm, 330 nm, and 400 nm. At $h$ = 330 nm (twist angle of 180 °), the chiral asymmetry factor at the resonance reaches $g$ = 1.25 (Figure S2f). This results from the interference of a higher-order mode at 598 nm (Figure S2b), which elevates the extinction response under RCP excitation (Figure S2e) and thus prevents the asymmetry factor from significantly approaching the theoretical limit

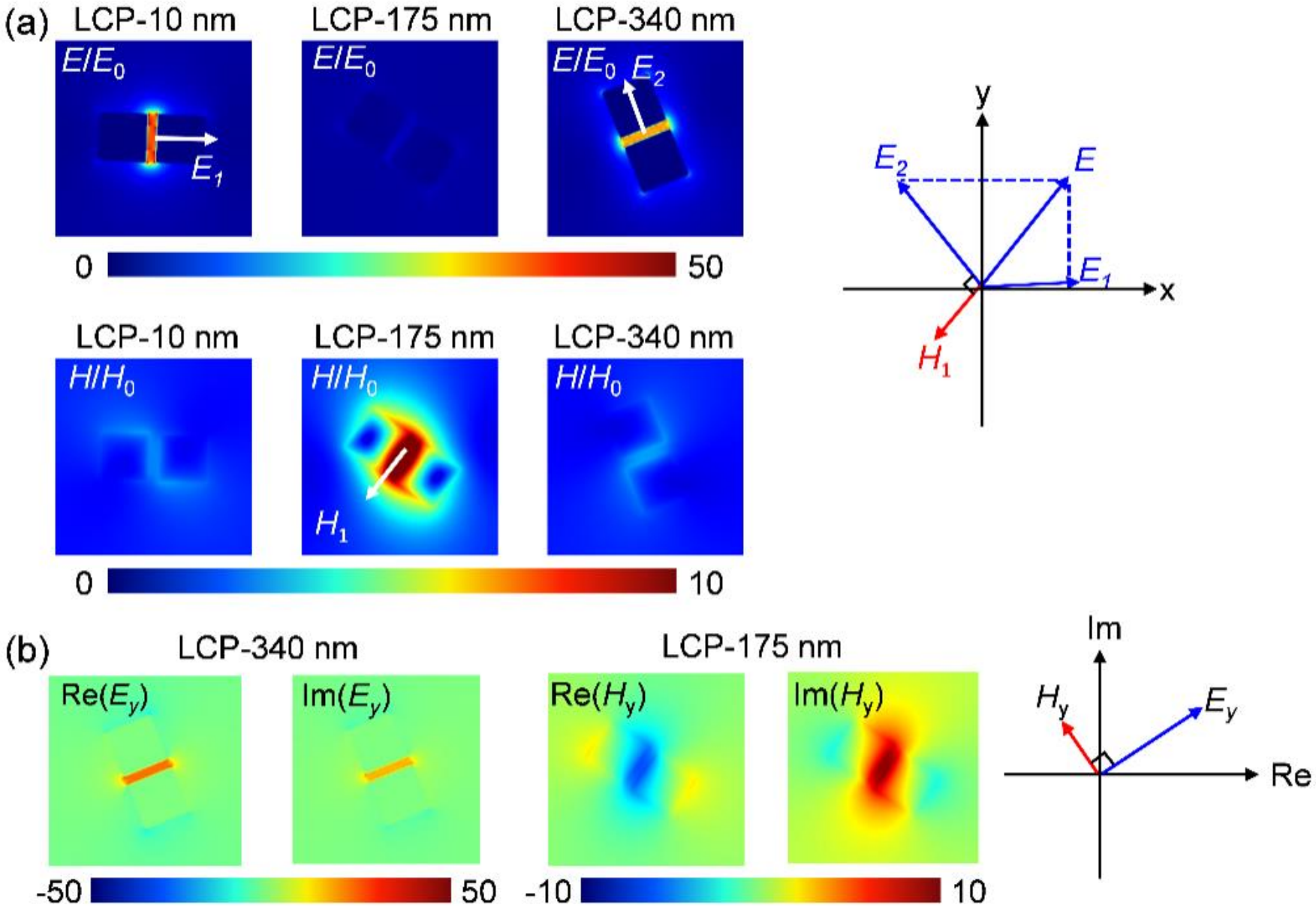

**Figure S3.** Near-field and phase in a twisted dimer structure under LCP. (a) Electric and magnetic near-field distributions at heights $Z$ = 10 nm,175 nm, and 340 nm for the twisted dimer structure under LCP excitation. The white arrows in (a) indicate the directions of the electric and magnetic near-fields respectively. The resultant electric fields of $E_1$ at Z = 10 nm and $E_2$ at Z = 340 nm are antiparallel to the magnetic field $H_1$ at Z = 175 nm. (b) Real and imaginary parts of $E_y$ at $Z$ = 340 nm and $H_y$ at $Z$ = 175 nm, where $Ey$ and $H_y$ are perpendicular to each other.

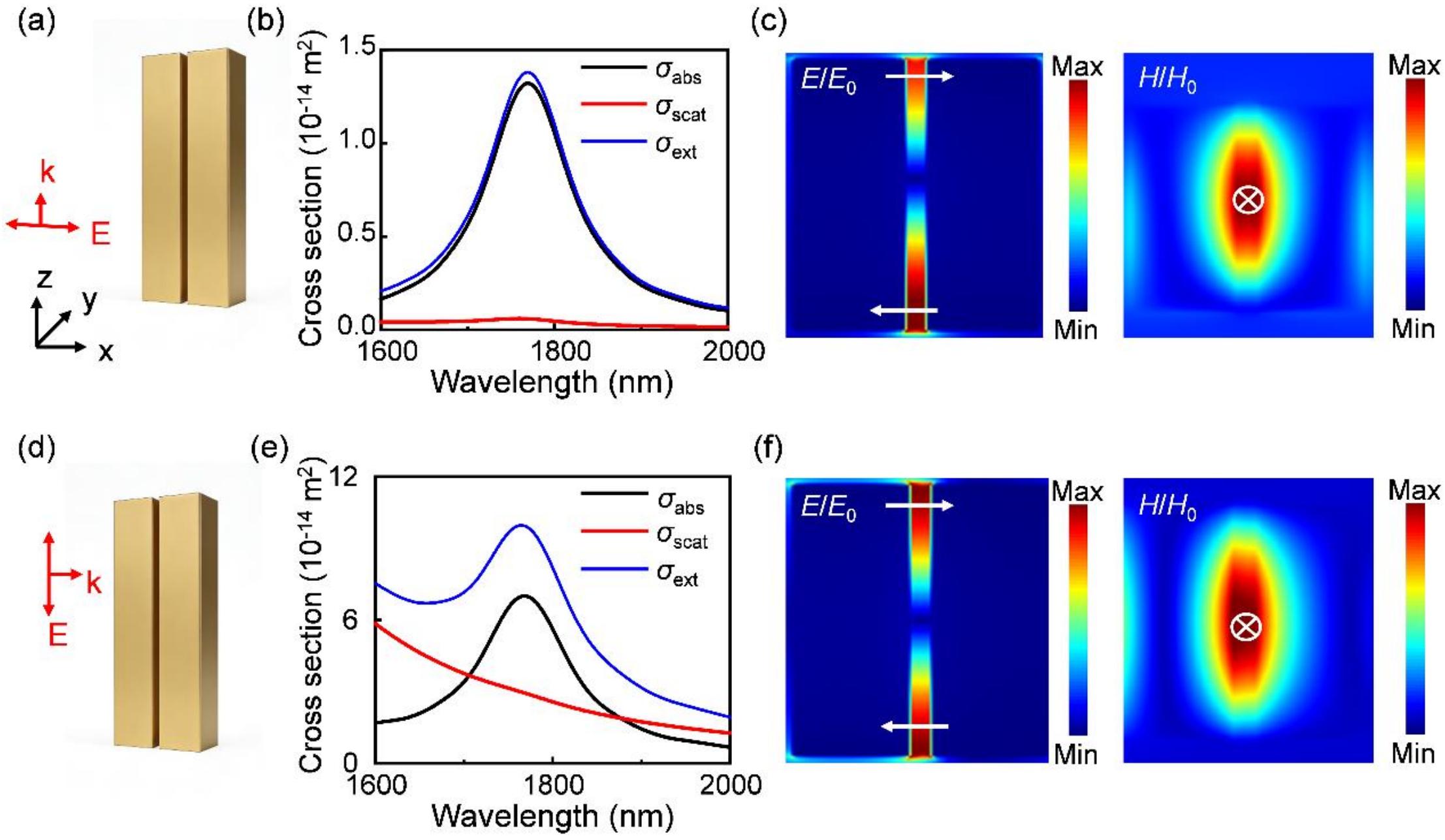

**Figure S4.** Optical response of a coupled nanorod dimer to *x*- and *z*-polarized light. (a) Schematic diagram of the coupled nanorod dimer under *x*-polarized plane wave excitation. The length, width, and height of each nanorod are 50 nm, 50 nm, 350 nm, and the gap between the two nanorods is 10 nm. (b) Corresponding absorption (black), scattering (red), and extinction (blue) cross-sections for *x*-polarized excitation. (c) Electric and magnetic near-field distributions at the resonance wavelength of 1831 nm for *x*-polarized excitation. The white arrows indicate the directions of the electric and magnetic fields. (d–f) Corresponding results under *z*-polarized plane wave excitation.

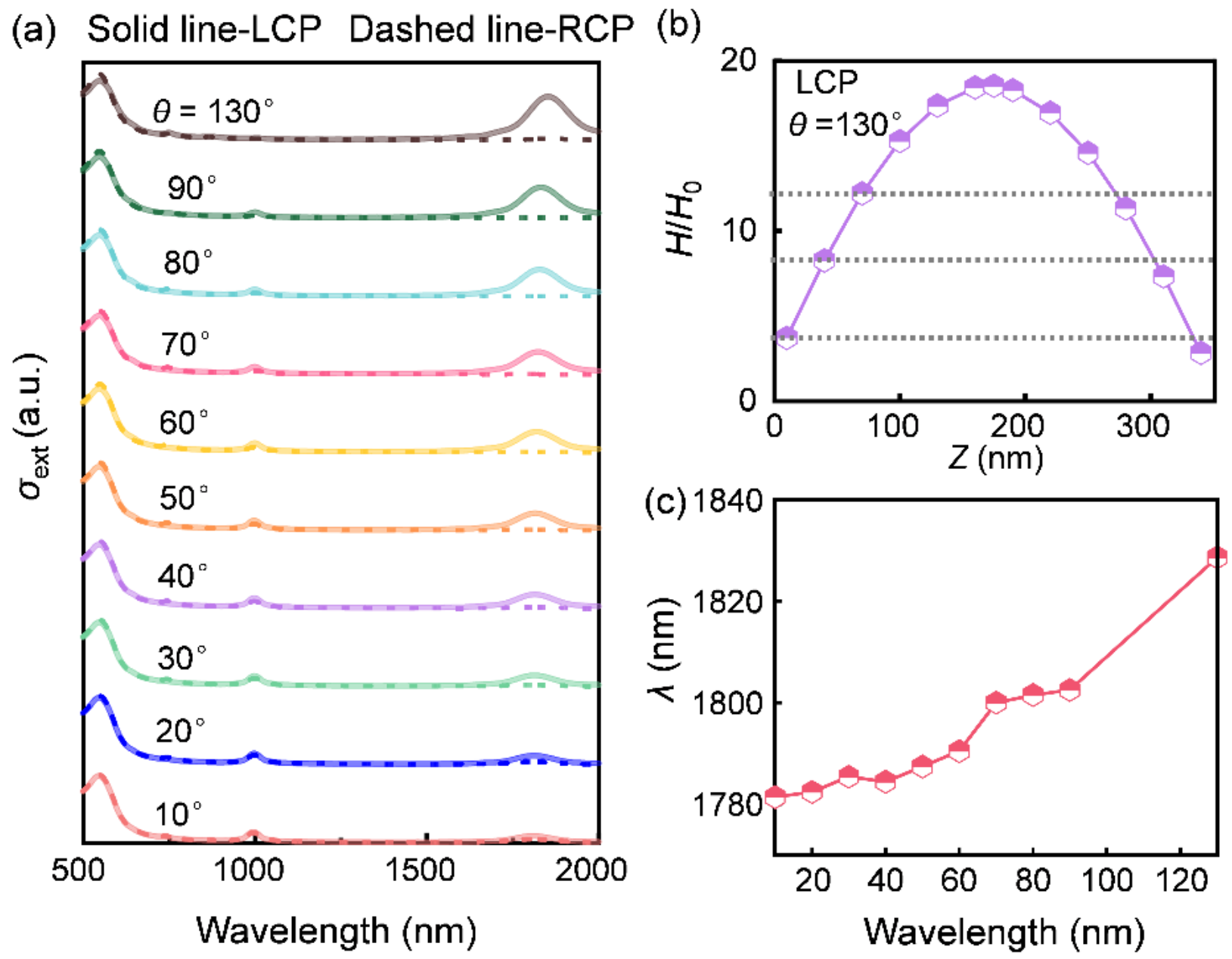

**Figure S5.** Influence of the twist angle $\theta$. (a) Extinction spectra under LCP (solid curves) and RCP (dashed curves) excitation for the twisted dimer structure with equal lengths and widths $L_1 = L_2 = 50$ nm, $h = 350$ nm, and $G = 10$ nm, as the helical twist angle $\theta$ is varied from 10° to 130°. (b) Magnetic field enhancement at the center of the twisted dimer structure with the twist angle of 130° as a function of height $Z$. The gray dashed line indicates that the field enhancements at two heights symmetric about the central height $Z = 175$ nm (e.g., $Z = 10$ nm and $Z = 340$ nm) are asymmetric. (c) Corresponding resonance wavelengths as functions of the twist angle.

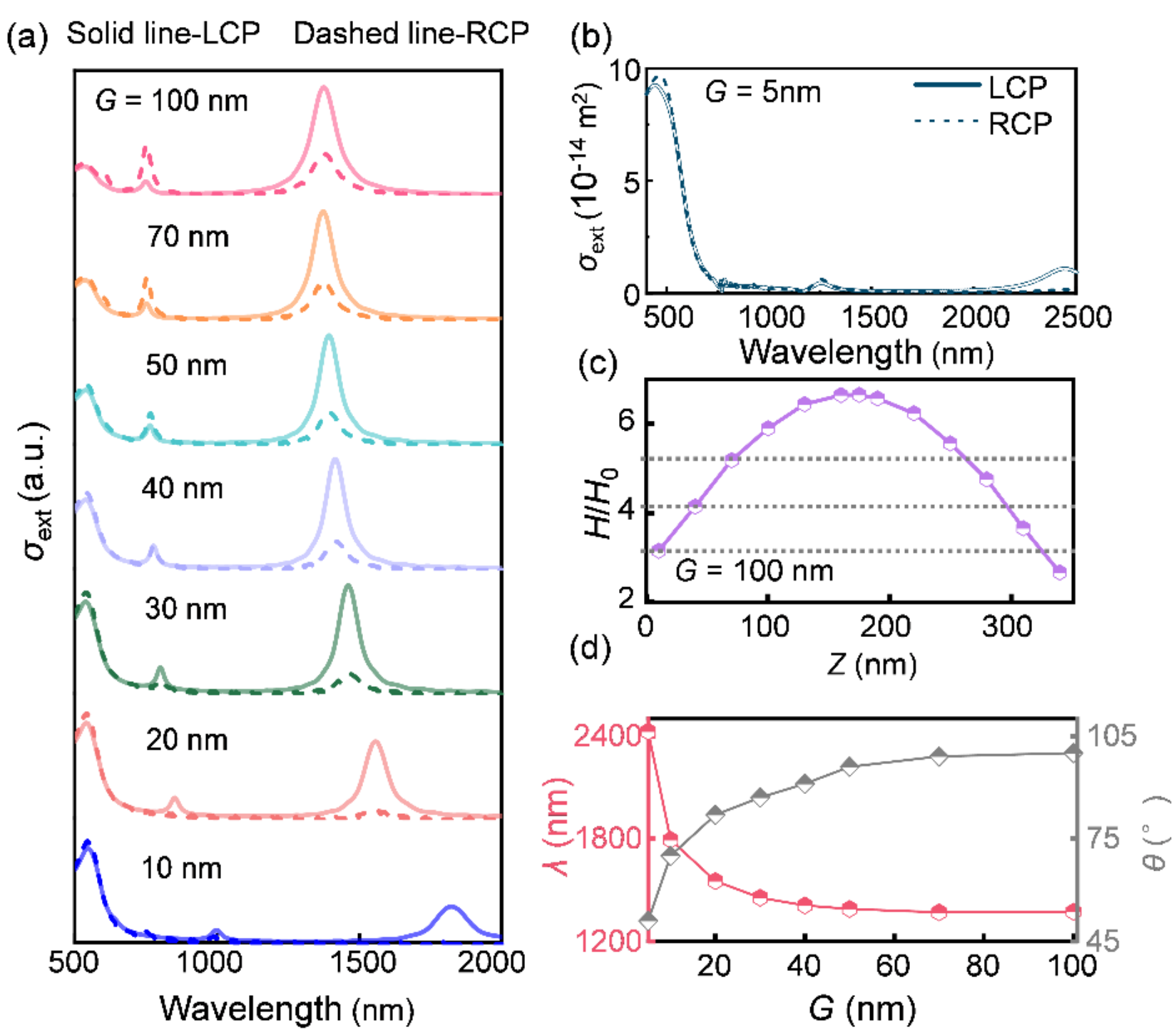

**Figure S6.** Influence of the gap $G$. (a) Extinction spectra under LCP and RCP excitation as the gap $G$ is increased from 10 nm to 100 nm, with the twisted dimer structure having equal lengths and widths $L_1 = L_2 = 50$ nm, height $h = 350$ nm, and the twist angle $\theta$ fixed at the optimal twist angle. (b) Extinction spectra under LCP (solid curve) and RCP (dashed curve) excitation for $G = 5$ nm. (c) Magnetic field enhancement at the center of the twisted dimer structure with the gap of 100 nm as a function of height $Z$. The gray dashed line indicates that the field enhancements at two heights symmetric about the central height $Z = 175$ nm (e.g., $Z = 10$ nm and $Z = 340$ nm) are asymmetric. (d) Resonance wavelength and optimal twist angle as functions of the gap size.

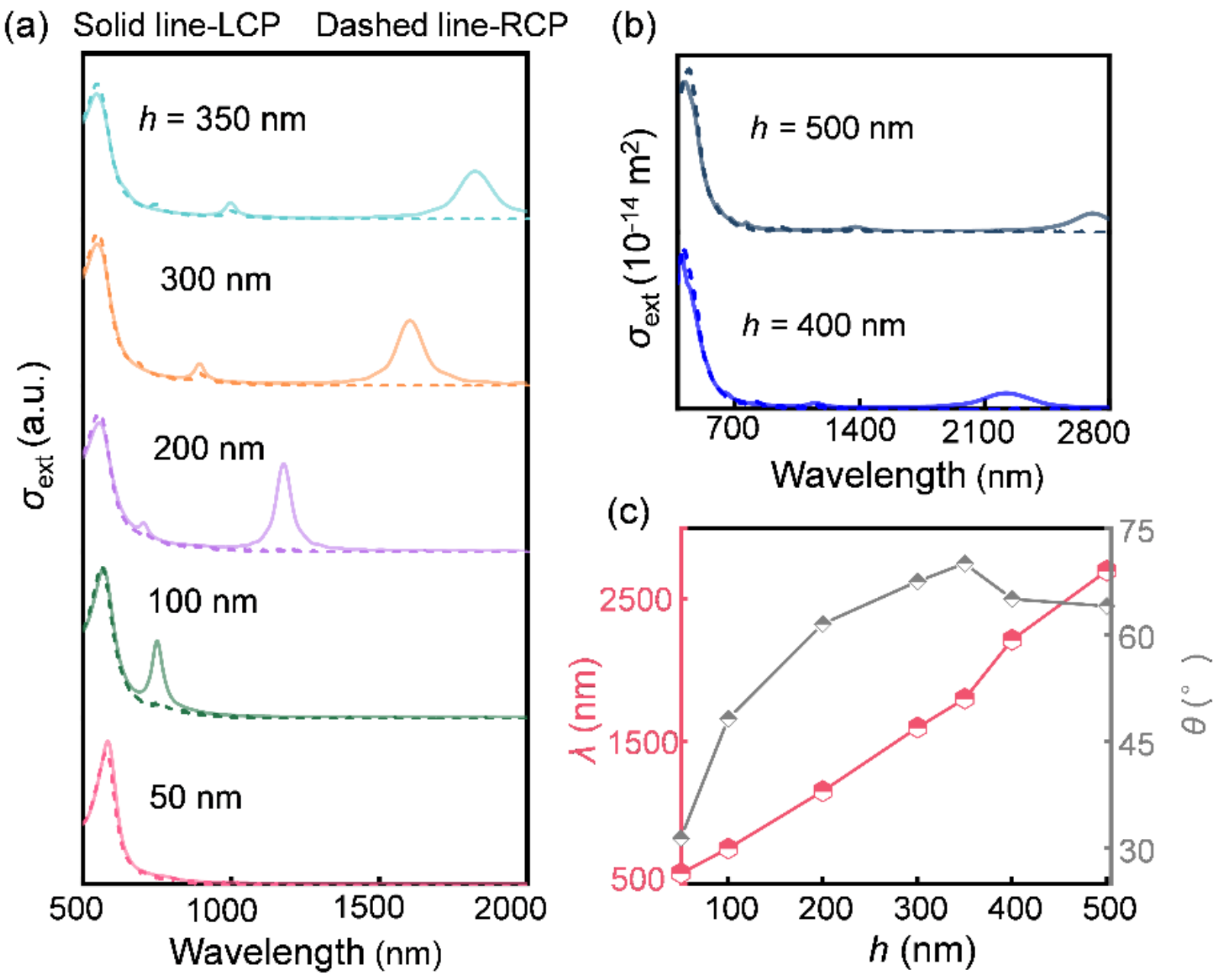

**Figure S7.** Influence of the height $h$. (a) Extinction spectra under LCP and RCP excitation as the height $h$ is increased from 50 nm to 350 nm, with the structure having equal lengths and widths $L_1$ = $L_2$ = 50 nm, $G$ = 10 nm, and the twist angle $\theta$ fixed at the optimal twist angle. (b) Extinction spectra under LCP (solid curve) and RCP (dashed curve) excitation for $h$ = 400nm and 500 nm. (c) Resonance wavelength and optimal twist angle as functions of the gap size.

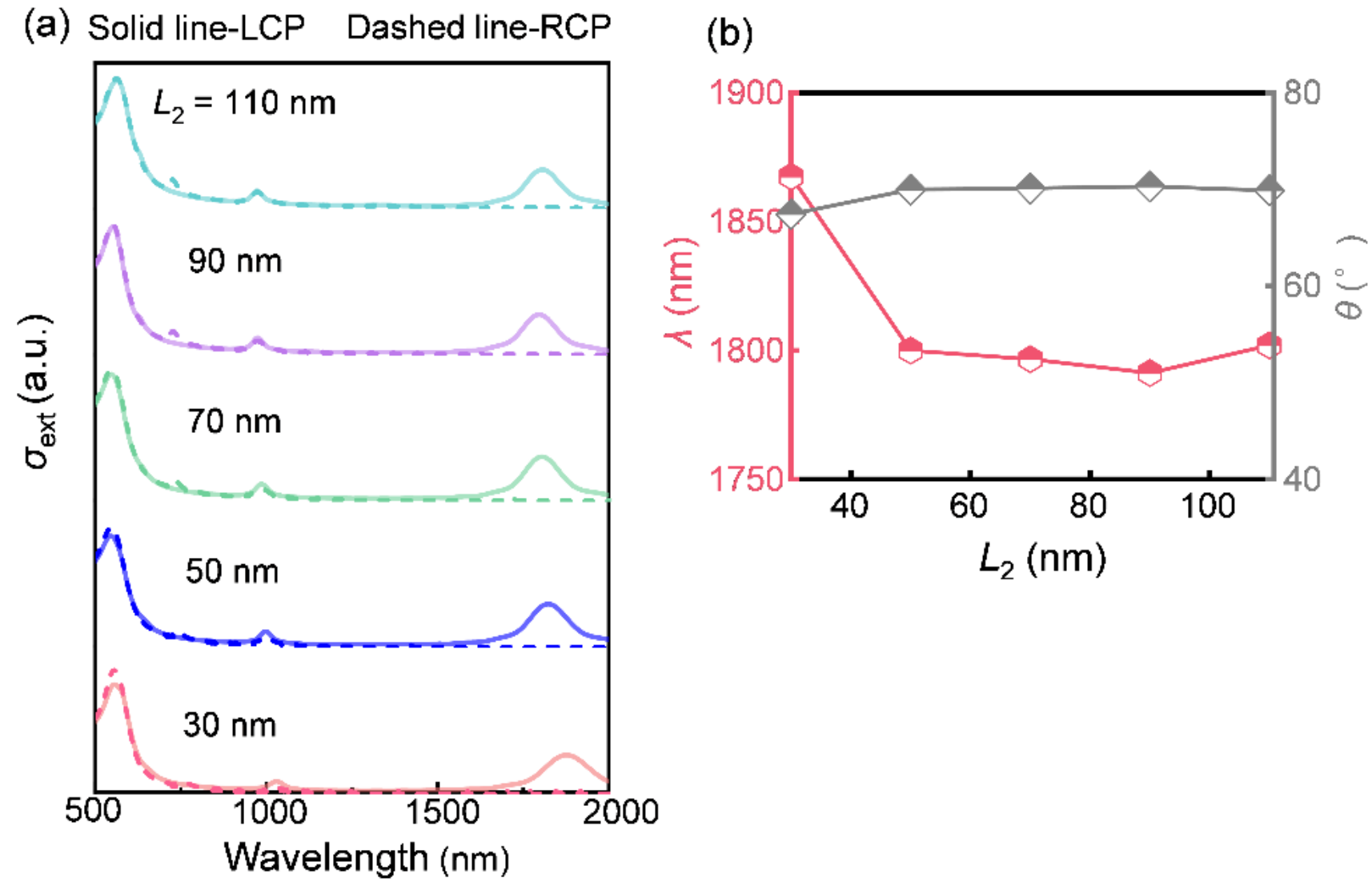

**Figure S8.** Influence of the width $L_2$. (a) Extinction spectra under LCP and RCP excitation as the width $L_2$ is increased from 30 nm to 110 nm, with the structure length $L_1$ = 50 nm, $G$ = 10 nm, $h$ = 350nm, and the twist angle $\theta$ fixed at the optimal twist angle. (b) Resonance wavelength and optimal twist angle as functions of the width.

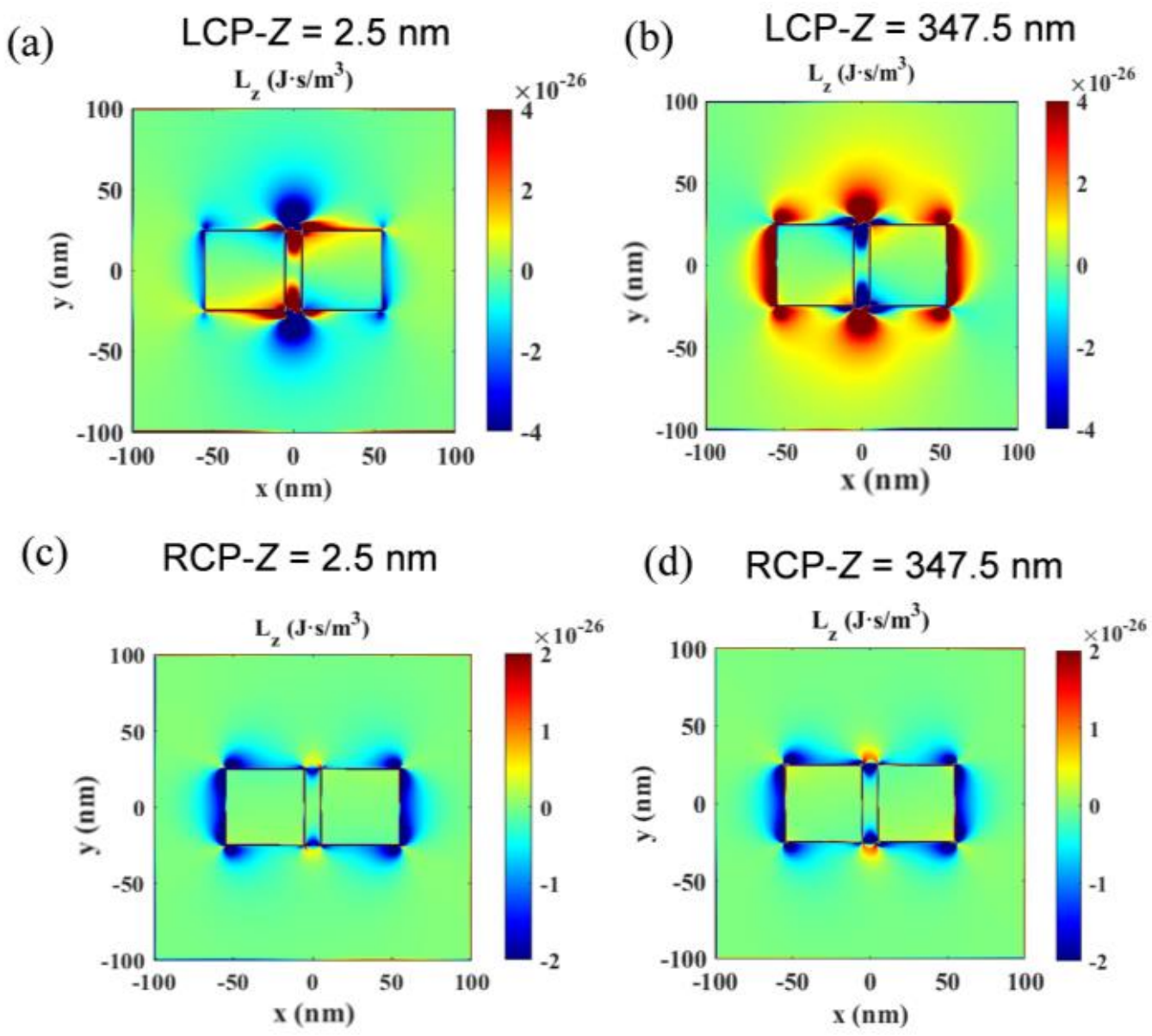

**Figure S9.** *L*z distributions under circular polarization. (a, b) Distributions of the orbital angular momentum density $L_z$ under LCP excitation at heights $Z$ = 2.5 nm (a) and $Z$ = 347.5 nm (b), respectively. (c,d) Distributions of the orbital angular momentum density $L_z$ under RCP excitation at heights $Z$ = 2.5 nm (c) and $Z$ = 347.5 nm (d), respectively.

When calculating the spin angular momentum and orbital angular momentum, unreasonable points appeared in the electric and magnetic fields at the edges of the structure due to the mesh discretization in FDTD. To address this issue, we rotated the structure around the $Z$-axis by an angle $\theta = (Z / \lambda) \times 360°$ to keep the calculation plane flat (i.e., with no relative rotation). For example, at $Z$ = 2.5 nm and $Z$ = 347.5 nm, the structure was rotated by 0.5° and 69.5°, respectively, thereby avoiding the generation of unreasonable points at the structural edges caused by mesh artifacts. However, when calculating the orbital angular momentum shown in Figure 4, unreasonable points still appeared despite the aforementioned rotation to maintain flatness, and we removed these points.